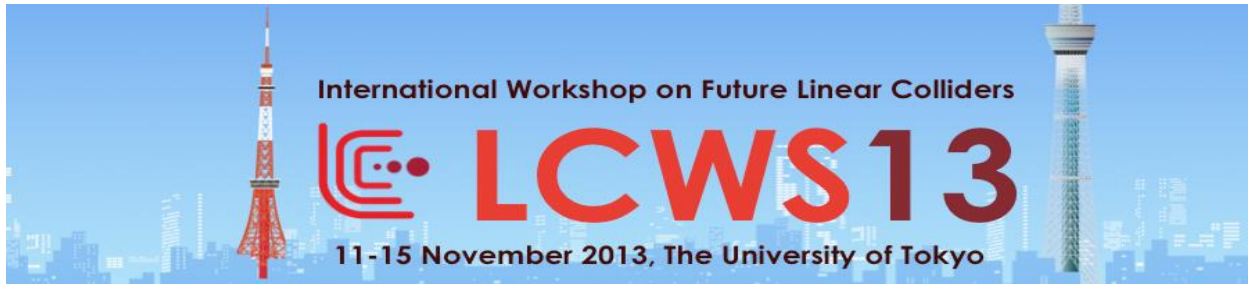

# Recent DHCAL Developments

**José Repond**[1]
*Argonne National Laboratory, Argonne, IL 60439*
E-mail: repond@anl.gov

**On behalf of the CALICE collaboration**

**Abstract:** This talk reports on recent progress concerning the development of a Digital Hadron Calorimeter with Resistive Plate Chambers as active elements. After the successful testing of a Digital Hadron Calorimeter prototype – the DHCAL - in the Fermilab and CERN test beams, the DHCAL group is now tackling some of the remaining technical issues which were not addressed specifically with the prototype. The talk reports on developments related to the RPC chamber design, to improvements in the RPC rate capabilities, the high voltage distribution system, and a gas recirculation system.

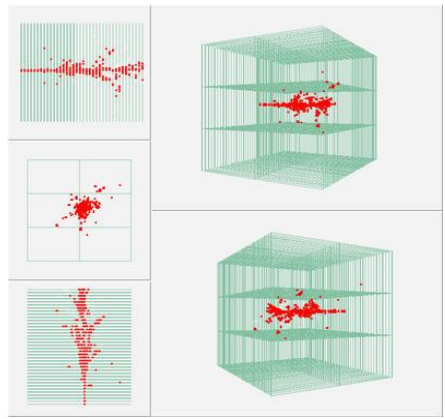

**Talk presented at the International Workshop on Future Linear Colliders (LCWS13)**
**Tokyo, Japan, 11-15 November 2013**

---

[1] Speaker

1. **Introduction: the DHCAL**

The Digital Hadron Calorimeter or DHCAL is a large scale prototype of an imaging hadron calorimeter using Resistive Plate Chambers (RPCs) as active elements. In past test beam campaigns, the DHCAL contained up to 54 1 x 1 $m^2$ layers interleaved with passive absorber material. The readout was embedded into the calorimeter (for the first time in the history of calorimeters) and was segmented into 1 x 1 $cm^2$ pads, each read out with a 1-bit (or digital) resolution. The device counted close to 500,000 readout channels, currently a world record for calorimetry and RPC-based systems.

In the test beams, the DHCAL layers were inserted into a main stack of 38 or 39 layers, followed by a tail catcher with up to 15 layers. For the tests performed at Fermilab, the main stack contained steel absorber plates. At CERN the absorber plates contained a Tungsten based alloy. In both cases, the tail catcher featured steel absorber plates.

In the various test beam campaigns combined, spanning the years 2010 – 2012, the DHCAL recorded of the order of 14 Million muon events and 36 Million secondary beam events, where the latter contained a mixture of electrons, muons, pions and protons.

With the completion of the test beam activities, the emphasis of the DHCAL group returns to the technical issues not specifically addressed by the prototype. This talk reviewed the current, major R&D activities. For measurements and results obtained with the DHCAL prototype in the Fermilab and CERN test beams, see the other DHCAL talks [1-3] presented at this conference.

2. **1-glass RPC design**

Traditionally RPCs feature two glass-plates with the gas volume sandwiched between them. Recently, the Argonne group proposed a novel design of RPCs, which omits the second glass plate and uses the readout board on the anode side to define the gas volume. As a consequence, the readout pads are located directly in the gas volume, thus reducing the distance between the avalanches and their electronic pick-up. Figure 1 shows a photograph of such a chamber with the approximate dimensions of 32 × 48 $cm^2$. To date several 1-glass RPCs prototypes have been assembled and have been tested successfully with comic rays.

The 1-glass RPCs offer a number of distinct advantages:

- The overall thickness of the chambers is reduced by the thickness of the second glass plate, typically of the order of 1 mm. If used as active elements, this reduces the overall thickness of a hadron calorimeter by up to 5 cm.
- The average pad multiplicity is close to unity and independent of the efficiency. Figure 2 shows a map of the efficiency and average pad multiplicity over the surface of the chamber. Note the uniform efficiency around 95% and the average pad multiplicity around 1.1. A uniform and stable average pad multiplicity close to unity significantly simplifies the calibration procedure aiming at equalizing the response of individual RPCs. In comparison, the chambers of the DHCAL showed an average pad multiplicity of about 1.65.
- The rate capability of RPCs is defined by the bulk resistivity of their resistive plates [4]. Eliminating one of the glass plates increases the rate capability by about a factor of two.
- Due to the geometrically narrower electric signals on the readout board, the position resolution of the chambers is vastly improved. Detailed measurements with a very finely segmented readout board, however, are still needed to establish the quantitative gain.

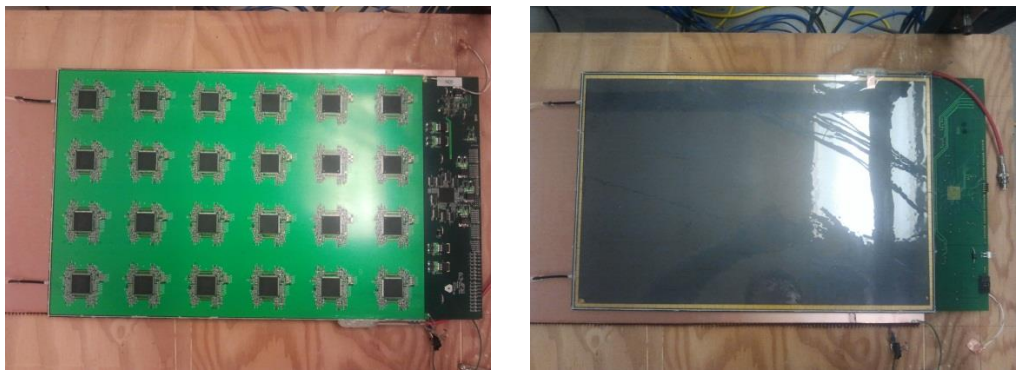

**Figure 1.** Photograph of the top (left) and the bottom (right) of a 1-glass RPC.

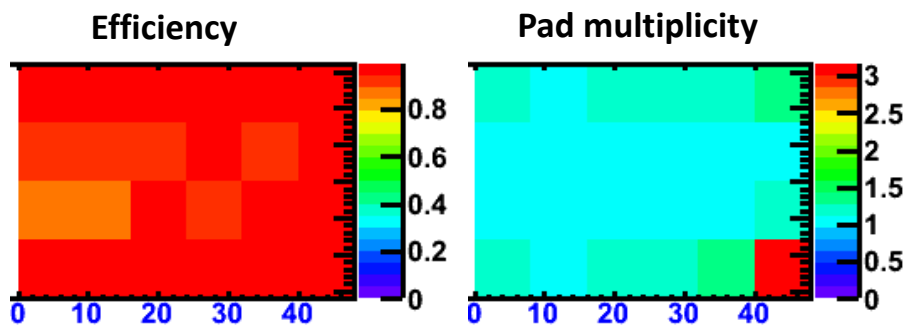

**Figure 2.** Maps of the efficiency (left) and average pad multiplicity (right) as measured with a 1-glass RPC and cosmic rays.

3. **High-rate RPCs**

The major weakness of RPCs is their limited rate capability. Assuming a constant flux of particles through the chambers, the efficiency is seen to decrease as a function of time. After several seconds the efficiency reaches a minimal value and remains at this lower value. This effect was clearly observed in tests of RPCs by B. Bilki et al. in the Fermilab test beam, see ref. [4] and Figure 3. An analytical model was developed [4] which successfully reproduced this behavior. In this model, the rate capability depends foremost on the bulk resistivity of the resistive plates. To improve the rate capability of RPCs, several groups worldwide currently are developing materials with bulk resistivities in the range of $10^8$ – $10^{10}$ $\Omega$cm.

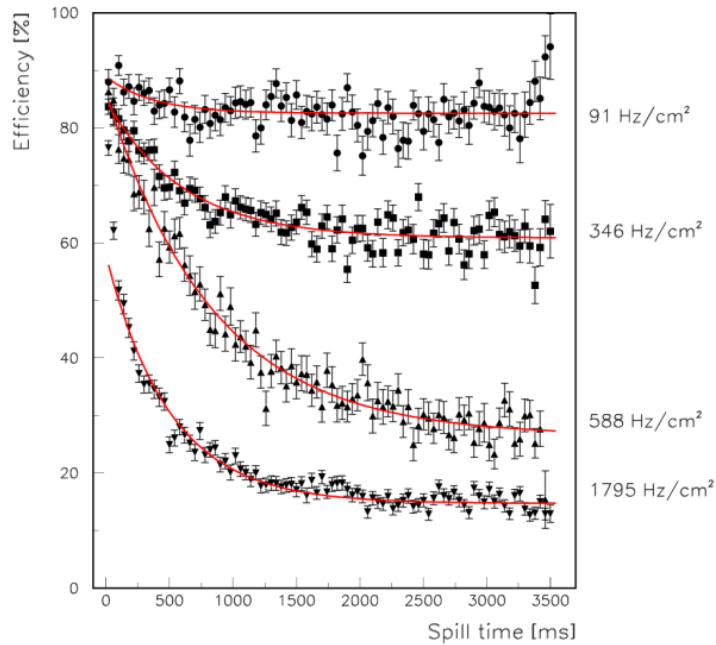

**Figure 3.** Efficiency for detecting minimum ionizing particles as function of the time since the beginning of a particle spill in the Fermilab test beam and for different beam intensities. The red lines are the results of calculations based on an analytical model. See ref. [4] for more details.

The DHCAL group pursues three different approaches to increase the rate capability of RPCs:

1) *Development of Bakelite with lower bulk resistivity.* This development is being carried out in co-operation with University of Science and Technology of China (USTC) and University of Michigan. The new Bakelite is being produced in China in cooperation with the above groups. Several chambers have already been assembled and appear to perform as expected. Detailed rate tests are still to be performed.
2) *Development of Bakelite with embedded resistive layer.* This development is also being carried out in co-operation with USTC and University of Michigan. Typically the resistive layer is applied on the outer surface of the resistive plates. Bakelite being a layered material offers the option to move the resistive layers inside the plate itself, as illustrated in Figure 4. The result is a reduced effective bulk resistivity of the plates, by up to an order of magnitude. Again, several chambers have already been assembled and appear to perform as expected. Detailed rate tests are still to be performed.
3) *Development of semi-conductive glass.* This development is being carried out in co-operation with COE College (Iowa) and University of Iowa. COE College is developing a Vanadium based glass with tunable bulk resistivity, which depends on the concentration of Vanadium. Small plates were produced with a bulk resistivity of approximately $10^8$ $\Omega$cm and were used to assemble an RPC. In the meantime, larger plates, see Figure 5, are being produced with improved properties. A first RPC using these plates will be assembled shortly.

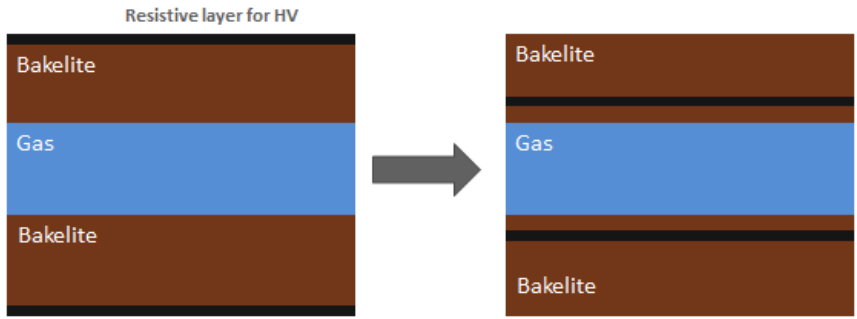

**Figure 4.** Cross section through a typical Bakelite RPC (left) and cross section through a Bakelite RPC with embedded resistive layer (right).

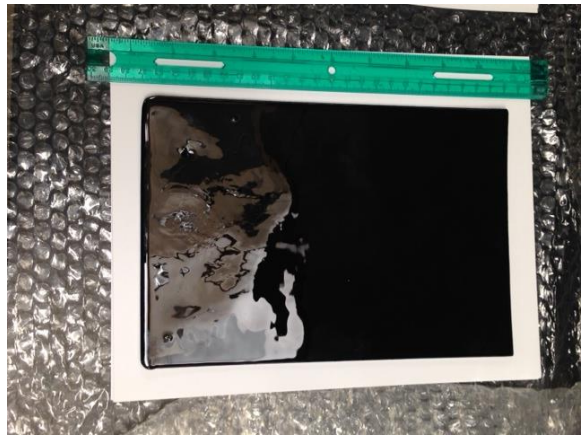

**Figure 5.** Photograph of a semi-conductive glass plate as produced by COE College (Iowa). The bulk resistivity was measured to be approximately $10^{10}$ $\Omega$cm.

### 4. High-voltage distribution system

Imaging calorimeters based on gaseous detectors need to provide high voltage to every single layer of a module. Routing cables from the power supplies directly to each individual layer appears to be prohibitively expensive and unacceptable from the point of view of the dead material budget. Therefore, a distribution system is needed which utilizes a single power supply and can

1) Distribute the high voltage to a larger number (of the order of 50) of layers,
2) Turn on and off layers individually,
3) Tune the high voltage of individual layers by several 100 V up or down, and
4) Monitor both the current and the voltage of each layer.

Such a system is being developed by University of Iowa. A first board with a single channel has been assembled and performed satisfactorily with RPCs. Using this distribution system with RPCs did not alter the accidental hit rates. Turning on and off the (single) channel did not trip the high voltage supply.

Unfortunately, due to lack of funding this activity is momentarily stopped.

## 5. Gas recycling system

The default gas used for RPCs operating in avalanche mode is a mixture of green house gases, see Table I. In the past, the exhaust from RPC was often simply vented into the atmosphere. Naturally, for larger systems or for prolonged operation this way of disposing of the exhaust is not acceptable from neither the financial nor the environmental point of view.

**Table I.** Typical gases used in RPCs operated in avalanche mode and their global warming potential (GWP).

| Gas | Fraction [%] | GWP (100 years, $CO_2$ = 1) | Fraction $\times$ GWP |
|---|---|---|---|
| Freon 134a | 94.5 | 1,430 | 1351 |
| Isobutan | 5.0 | 3 | 0.15 |
| $SF_6$ | 0.5 | 22,800 | 114 |

Under the leadership of University of Iowa, the DHCAL group has initiated the development of a gas recirculation/recovery system. The system preserves the atmospheric pressure at the exhaust of the chambers and is thus dubbed 'Zero Pressure Containment' system. Figure 6 shows a rough sketch of the system, which involves six separate tanks. A prototype of the system is currently being assembled and will be tested with the RPCs of the DHCAL.

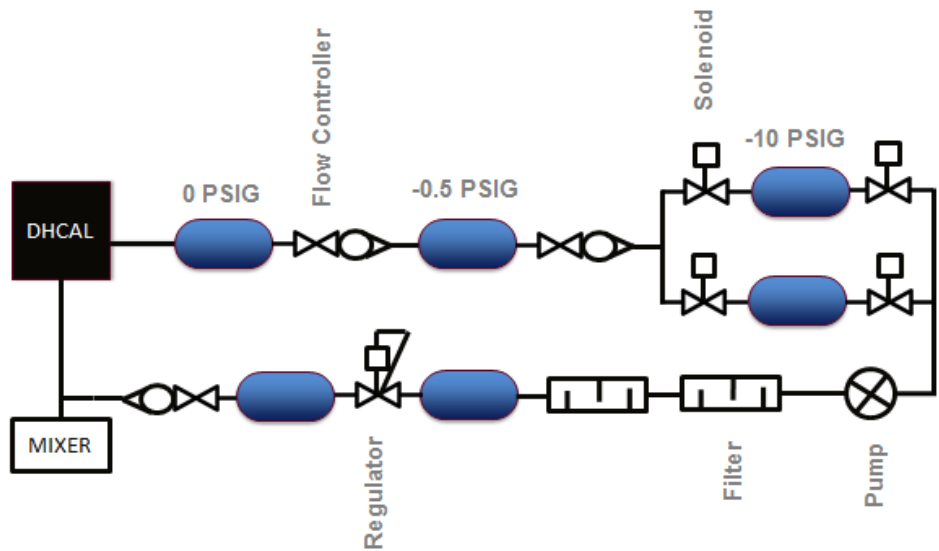

**Figure 6.** Sketch of the 'Zero Pressure Containment' system to recycle the RPC gas.

## 6. Summary

After successfully testing the DHCAL concept in the Fermilab and CERN test beams, the group is tackling some of the remaining technical issues before considering the technology of an RPC-based imaging calorimeter mature enough to be proposed for a detector at a future lepton collider. The group is developing 1-glass RPCs, low-resistivity Bakelite, semi-conductive glass, a high-voltage distribution system and a gas recirculation system.

## Acknowledgment

The author wishes to thank the organizers for a wonderful conference, boasting an exciting scientific program, as well as an exceptional and memorable social dinner.